\documentclass[preprintnumbers, showpacs, amsmath, aps, amsmath, 12pt]{revtex4}
\usepackage[dvips]{graphicx}
\usepackage{dcolumn}\begin{Large}\begin{tiny}\end{tiny}\end{Large}
\usepackage{bm}
\begin{document}
\title{Effects of dipolar interactions on the zero-field-cooled magnetization of a nanoparticle assembly}
\author{M. Azeggagh}
\email{azeggagh@physique.uvsq.fr}
\author{H. Kachkachi}
\affiliation{Groupe d'Etude de la Mati\`ere Condens\'ee, Universit\'e de Versailles St. Quentin, CNRS UMR8635, 45 av. des Etats-Unis, 78035 Versailles, France}
\date{\today}
\begin{abstract}
We investigate the effect of (weak) dipolar interactions on the field behavior of the temperature at the maximum of the zero-field-cooled magnetization of a polydisperse assembly of nanoparticles.
For this purpose we extend the Gittleman-Abeles-Bozowski model for the zero-field-cooled magnetization by computing the contribution of dipolar interactions to the longitudinal relaxation time.
We show, in good qualitative agreement with many experimental observations, that the temperature at the maximum of the zero-field-cooled magnetization as a function of the applied field changes from a bell-like to a monotonically decreasing curve when the intensity of the dipolar interactions, or equivalently the sample concentration, increases.
\end{abstract}
\maketitle
\section{Introduction}
One of the direct technological applications of magnetic nanoparticles is magnetic recording. The storage capacity of the media can be considerably increased by devising denser assemblies of smaller and smaller particles. However, this brings a dilemma because small particles become superparamagnetic, i.e., thermally unstable, well below the room temperature. Moreover, high density in assemblies entails strong dipole-dipole interactions (DDI) between the particles, and in technological applications such as magnetic recording, this is an issue of special importance since DDI have been widely recognized as being responsible for the deterioration of the signal-to-noise ratio (see e.g., Refs.~\onlinecite{sharrock90ieee, johnson91jap} and references therein). As such, an optimal material, with appropriate anisotropy and other physical parameters, has still to be devised.
On the other hand, the study of nanoparticle assemblies brings new hurdles to theorists, at least, since one is faced with tremendous difficulties related with DDI between particles, together with the distributions of volume and anisotropy axes.
In spite of that, DDI in nanoparticle assemblies has triggered much interest due to many new phenomena that emerge from the collective behavior of the particles and also because these interactions have always constituted a challenging issue in many areas of physics.

To analyze and eventually understand experiments on the dynamics of interacting nanoparticle assemblies, and in particular to understand the dynamical response, such as the ac susceptibility and zero-field-cooled (ZFC) and field-cooled (FC) magnetizations, one needs to know how DDI affect the switching process and the relaxation time of the nanoparticles.
While a fair understanding of the mechanisms underlying the ZFC-FC magnetization process and  ac susceptibility has been achieved in the case of non-interacting assemblies, many experimental results on interacting assemblies remain unexplained.
Obviously, this is mainly due to the long range of DDI and also to the complexities of the very calculation of the relaxation time itself.
Recently, J\"onsson and Garcia-Palacios \cite{jongar01epl} [see also \cite{garpalgar04prb}] obtained an approximate expression for the relaxation rate of a weakly interacting monodisperse assembly of macropins with textured or randomly-distributed anisotropy. The macrospin approximation here means that a particle is represented by a macroscopic magnetic moment, i.e., ignoring its internal structure. This picture of the particle will henceforth be referred to as the one-spin problem (OSP). In the literature, this is also known as the coherent-rotation limit.
Then using the simple Debye relaxation model they investigated the effect of DDI on the ac susceptibility and in particular the displacement of the maximum of its real and imaginary components.
In their explanation of this effect they emphasized the important role played by damping in the relaxation processes in the presence of a transverse field in addition to the effect of the change in the energy barriers, that was commonly believed to play the major role. The role of a transverse field is played here by the transverse component of the dipolar field.

Experimental results obtained for ferrofluids \cite{luoetal91prl} and later for $\gamma$-Fe$_{2}$O$_{3}$ nanoparticles \cite{sapetal97prb} indicated that for dilute samples (weak DDI), the temperature $T_{\max}$ at the maximum of the ZFC magnetization first increases with increasing field, attains a maximum and then decreases. More experiments performed on the $\gamma$-Fe$_{2}$O$_{3}$ particles dispersed in a polymer \cite{ezzir98phd, kacetal00jpcm} matrix confirmed the previous results for dilute samples and showed that, on the contrary, for concentrated samples (strong DDI) $T_{\max }$ is a monotonically decreasing function of the magnetic field.
The shift of this maximum was also studied with different techniques in various types of nanoparticles, see for instance \cite{mamnak98jmmm, mamiyaetal02prl, hansonetal95jcmp}.
In Ref.~\onlinecite{kacetal00jpcm} it was shown that the bell-like shape of $T_{\max}(H)$ is not very sensitive to the intrinsic properties of the particles, of course in the OSP approximation.
Exact numerical calculations \cite{geoetal97acp, coffeyetal98prb, kennedy97phd} of the smallest eigenvalue of the Fokker-Planck operator invariably led to a monotonic decrease in the blocking temperature, and thereby in the temperature $T_{\max }$, as a function of the magnetic field.
Indeed, it was shown that the expression of the single-particle relaxation time does not play a crucial role and that even the (relatively) simple N\'eel-Brown expression for the relaxation time in a longitudinal field leads to a maximum in $T_{\max }(H)$. What seemed to play a crucial role is the fact that the magnetization, formulated within the Gittleman, Abeles, and Bozowski (GAB) model \cite{gitetal74prb}, has a superparamagnetic contribution that is a non-linear function (such as Langevin's) of the magnetic field. The magneto-crystalline anisotropy and the volume-distribution width also have strong influence.
The issue of the effect of DDI on $T_{\max }(H)$, namely the disappearance of the maximum when the intensity of interactions increases, was left open in Ref.~\onlinecite{kacetal00jpcm}.
In the present work we revisit this issue after generalizing the work of Ref.~\onlinecite{jongar01epl} to include the static magnetic field and magnetic-moment distribution (polydisperse assembly) of nanoparticles in the OSP approximation.
We then investigate the effect of (weak) DDI on the ZFC magnetization and in particular on $T_{\max }(H)$.

The present work is organized as follows:
After fixing the notation and defining the model Hamiltonian, we explain our formalism for computing the ZFC magnetization within the GAB model: we compute the contribution of DDI to the longitudinal relaxation rate, and then show how the GAB model is accordingly extended.
Next, we discuss the effect of DDI on $T_{\max }(H)$ of a given assembly and compare the results for two materials, namely maghemite and cobalt.
\section{\label{basics}Notation and basic formulas}
We consider an assembly of magnetic moments ${\bf m}_i = m_i{\bf s}_i,\, i=1,\ldots,{\cal N}$ of magnitude $m_i$ and direction ${\bf s}_i$, with $\vert{\bf s}_i\vert=1$. The magnitude of the magnetic moment ${\bf m}_i$ is given in terms of the Bohr magneton $\mu_B$, i.e., $m_i=n_i\mu_B$, and the numbers $n_i$ are log-normal distributed.
Each magnetic moment is assigned a uniaxial easy axis ${\bf e}_i$, and for the assembly these axes are randomly distributed.
The energy of a magnetic moment ${\bf m}_i$ with uniaxial anisotropy axis ${\bf e}_{i}$, interacting with all the others via DDI, in the magnetic field ${\bf H}=H{\bf e}_h$, reads [after multiplying by $-\beta=-1/k_BT$],
\begin{equation}\label{Ei_dimless}
{\cal E}_i = x_i\left({\bf s}_i\cdot{\bf e}_{h}\right) + \sigma_i({\bf s}_{i}\cdot{\bf e}_{i})^{2} + \xi_d\sum _{j<i}n_i n_j {\bf s}_{i}\cdot{\cal D}_{ij}\cdot{\bf s}_{j},
\end{equation}
where $x_i=xn_i, \sigma_i=\sigma_0 n_i$ with 
\begin{equation}\label{dimless_params}
x = \frac{\mu_BH}{k_BT}, \quad \sigma_0 = \frac{\mu_BK}{M_s k_BT}, \quad \xi_d = \left(\frac{\mu_0}{4\pi}\right)\frac{\mu_B^2/a^{3}}{k_{B}T},
\end{equation}
being the dimensionless energy parameters. Note that $\sigma_i=KV_i/(k_BT)$ (or simply $\sigma$ for a monodisperse assembly) is the commonly used notation for the reduced anisotropy-energy barrier height of the particle $i$.
${\cal D}$ is the DDI tensor defined as
\begin{eqnarray}\label{DDITensors}
{\cal D}_{ij} &\equiv& \frac{1}{r_{ij}^3}\left(3{\bf e}_{ij}{\bf e}_{ij}-1\right).
\end{eqnarray}
where ${\bf r}_{ij}={\bf r}_{i}-{\bf r}_{j}$, with ${\bf e}_{ij}={\bf r}_{ij}/r_{ij}$, is the vector joining the sites $i$ and $j$ and whose magnitude is measured in units of $a$, a characteristic distance on the matrix in which the particles are embedded.
More precisely, the parameter $a$ is taken as a real number times the mean diameter $D_m$ of the assembly, i.e., $a=k\times D_m$. Thus, large values of $k$ correspond to an isotropically inflated lattice with large distances between the magnetic moments, and thereby weak DDI.
\section{Zero-field-cooled magnetization and effects of dipolar interactions}
\subsection{Zero-field-cooled magnetization}
The dynamic response of the OSP assembly is given by the ac susceptibility. For a given particle with an arbitrary angle $\psi$ between its anisotropy easy axis and the field direction, the effective susceptibility may be written as
\[
\chi(\omega) = \chi_\parallel\cos^2\psi + \chi_\perp\sin^2\psi. 
\]
Shliomis and Stepanov \cite{shlste93jmmm} proposed a simple Debye form for $\chi(\omega)$, which can be generalized to describe the effect of a longitudinal bias field by writing
\begin{equation}  \label{chi_SHS}
\chi_\mathrm{SHS} = \dfrac{\chi_\parallel(T,H)}{1+i\omega\tau_\parallel}\cos^2\psi + \dfrac{\chi_\perp(T,H)}{1+i\omega\tau_\perp}\sin^2\psi,
\end{equation}
where $\tau_\parallel$ and $\tau_\perp$ are appropriate longitudinal (interwell) and transverse (intrawell) relaxation times; $\chi_\parallel(T,H)$ and $\chi_\perp(T,H)$ are respectively the longitudinal and transverse components of the equilibrium susceptibility.

In the limit of a high anisotropy-energy barrier, i.e., $\sigma \gg 1,h=x/(2\sigma)\ll 1$, approximate expressions were found in Ref.~\onlinecite{jongar01prb} for the longitudinal and transverse components of the equilibrium susceptibility.
If one sets in (\ref{chi_SHS}) $\tau_\perp=0$ [instantaneous intrawell transverse response] and uses the high-barrier approximation one arrives at the GAB model \cite{gitetal74prb}, generalized to $H\neq 0$ and an arbitrary anisotropy-axis orientation. More precisely, upon evaluating the high-barrier expressions for $H = 0$, inserting the result in (\ref{chi_SHS}), setting $\tau_\perp=0$, and averaging over an assembly with randomly distributed anisotropy axes, one arrives at the expression proposed in Ref.~\onlinecite{gitetal74prb},
\begin{equation}  \label{chi_GAB_RA}
\chi_\mathrm{GAB} \simeq \chi_0\dfrac{1+\dfrac{i\omega\tau_\parallel}{\sigma}}{1+i\omega\tau_\parallel}\simeq \dfrac{\chi_0}{1+i\omega\tau_\parallel}.
\end{equation}

The real and imaginary components then read
\begin{equation}  \label{chi_GAB_RA_ReIm}
\chi^\prime = \dfrac{\chi_0 + \chi_1(\omega^2\tau_\parallel^2)}{1+\omega^2\tau_\parallel^2}, \quad \chi^{\prime\prime} = \dfrac{\omega\tau_\parallel\left(\chi_1-\chi_0\right) }{1+\omega^2\tau_\parallel^2}.
\end{equation}
where
\begin{equation}\label{Chi0Chi1_limits}
\chi_0 = \dfrac{M_s^2(T) V}{3 k_B T}, \quad \chi_1 = \dfrac{M_s^2(T) V}{3 K V}
\end{equation}
are respectively the susceptibility at thermodynamic equilibrium and the initial susceptibility of particles in the blocked state (see \cite{doretal97acp} and references therein).
Accordingly, starting from (\ref{chi_GAB_RA_ReIm}) the application of an alternating field yields: a) $\chi ^{\prime }=\chi _{0}$ if $\omega \tau \ll 1.$ At high temperature the magnetic moments orientate themselves on a great number of occasions during the time of a measurement, and thus the susceptibility is the superparamagnetic susceptibility $\chi _{0}.$ b) $\chi ^{\prime }=\chi _{1}$ if $\omega \tau \gg 1.$ At low temperature the energy supplied by the field is insufficient to reverse the magnetic moments the time of a measurement.
Then, the susceptibility is the static susceptibility $\chi _{1}.$ Between these two extrema there exists a maximum at the temperature $T_{\max }$. $\chi ^{\prime }$ can be calculated from (\ref{chi_GAB_RA_ReIm}) using the formula for the relaxation time $\tau $ appropriate to the anisotropy symmetry, and considering a particular volume $V$ one can determine the temperature $T_{\max }$.
Expression (\ref{chi_GAB_RA_ReIm}) of the dynamic susceptibility obtained for instantaneous transverse response is particularly suitable for the calculation of the ZFC magnetization and $T_{\max}$. Indeed, Eq.~(\ref{chi_GAB_RA_ReIm}) was used in Ref.~\onlinecite{kacetal00jpcm} to study the effect of anisotropy and volume distribution on $T_{\max}(H)$.
The formalism used can be summarized as follows.
In an assembly of particles with a volume distribution, $\chi^{\prime }$ can be calculated by postulating that at a given temperature and given measuring time, certain particles are in the superparamagnetic state and that the others are in the blocked state. The susceptibility is then given by the sum of two contributions \cite{gitetal74prb}
\begin{equation}\label{Susc5}
\chi ^{\prime }(T,\nu ) = \int\limits_{0}^{V_{c}}\mathcal{D}V\,\chi _{0}(T,V,\nu ) + \int\limits_{V_{c}}^{\infty }\mathcal{D}V\,\chi_{1}(T,V,\nu ),
\end{equation}
where $\mathcal{D}V$ is the measure of the log-normal volume distribution with parameters  $V_{0}$ and $\delta $
\begin{equation}\label{LNdist}
 \mathcal{D}V=\frac{1}{\delta \sqrt{2\pi }}\exp \left[ -\frac{\log ^{2}(\frac{V}{V_0})}{2\delta ^2}\right] \frac{dV}{V}.
\end{equation}
$V_{c}=V_{c}(T,H)$ is the critical volume defined as the volume for which $\tau_\parallel^{-1} = \nu_m$, where $\nu_m$ is the measuring frequency.
$V_{c}$ is the ``critical volume" that discriminates between the dominating populations of superparamagnetic particles of volume $V < V_c$ and blocked particles with $V >V_c$, and is experiment-dependent.

Eq.~(\ref{Susc5}) can be rewritten for the ZFC magnetization as follows
\begin{eqnarray}\label{MagSupMagBlock}
M_{zfc}(H,T,\psi) &=& \int\limits_{0}^{V_{c}}\mathcal{D}V\,\,M_{sp}(H,T,V,\psi) \\
&+& \int\limits_{V_{c}}^{\infty }\mathcal{D}V\,\,M_{b}(H,T,V,\psi ) \nonumber
\end{eqnarray}
where $M_{sp}=H\chi_0$ and $M_b=H\chi_1$ are the contributions to the magnetization from the superparamagnetic and blocked particles, respectively.

In the present work we extend this formalism to include (weak) DDI and to investigate their effect on $T_{\max}$.
For this purpose, it is necessary, in principle, to compute the contributions of DDI to both the equilibrium susceptibility in the numerator of Eq.~(\ref{chi_GAB_RA}) and to the relaxation time $\tau_\parallel$ in the denominator.
For the first calculation we can differentiate with respect to the applied field the expression obtained in Ref.~\onlinecite{kacaze05epjb} for the longitudinal magnetization taking account of DDI and anisotropy. We can also derive an expression for the transverse equilibrium susceptibility as a response to a transverse magnetic field \cite{garpal00acp}.
For the second calculation, we generalize the expression obtained in Ref.~\onlinecite{jongar01prb} for the relaxation time (including DDI) to include the volume distribution and the static magnetic field. This is done in the next section.
While the outcome of the first calculation is an insignificant quantitative modification of $T_{\max}$, the DDI contribution to the relaxation time yields a dramatic qualitative and quantitative effect since $T_{\max}$ changes from a bell-like to a monotonically decreasing function.
Indeed, with regard to Eq.~(\ref{MagSupMagBlock}), we will show that the change in the relaxation rate due to DDI induces a change in the critical volume $V_{c}$ and thereby a change in the dominating population of blocked or superparamagnetic particles.
\subsection{\label{sec:Gamma_DDI}Effect of DDI on the relaxation rate}
The idea is to introduce the (local) dipolar field ${\bf \xi}_i$ (note that ${\cal D}_{ii}=0$ and see notation in section \ref{basics}),
\begin{equation}\label{eff_field}
\xi_i^\mathrm{ddi}= \xi_d n_i\sum_j n_j{\cal D}_{ij}{\bf s}_{j}.
\end{equation}  
The relaxation rate of a magnetic moment that experiences this field can be computed using perturbation theory assuming that $\mid{\bf\xi}_i\mid\ll 1$ \cite{jongar01epl, garpalgar04prb}.
Accordingly, in Ref.~\onlinecite{luisetal04jpcm} an estimation of DDI was given for two samples of cobalt nanoparticles which indicates that the DDI field is of the order of $300$ Oe, which in reduced units, obtained after dividing by the corresponding anisotropy field of the order of $0.3$ T, is $\vert\xi/H_a\vert\sim 4\times10^{-3}-10^{-2}$. This is of course very small, which indeed suggests that in typical (relatively dilute) samples the above condition on the DDI field is often satisfied.
On the other hand, the magnetic field at which $T_{\max }(H)$ has a maximum at approximately $100$ Oe [see Fig. 1 of  Ref.~\onlinecite{kacetal00jpcm}, for maghemite particles], corresponding to a reduced field $h=H/H_a\simeq3\times 10^{-2}$.
This shows that even though $\xi$ is small it may still have a strong effect on $T_{\max }(H)$ because it is of the same order as the applied field in the relevant range.

The relaxation rate of a weakly DDI-interacting nanomagnet obtained in Ref.~\onlinecite{jongar01epl} is then written as
\begin{equation}\label{jongar_rr}
\Gamma_i \simeq  \Gamma_i^{(0)}\left[ 1+\frac{1}{2}\xi _{i,\parallel}^{2}+\frac{1}{4}F_{i}\xi _{i,\perp }^{2}\right] ,
\end{equation}
where $\Gamma_i^{(0)}$ is the relaxation rate of the nanomagnet at site $i$ in the absence of the DDI field $\xi_i^\mathrm{ddi}$ and is given by the (intermediate-to-high damping) N\'eel-Brown expression
\begin{widetext}
\begin{equation}\label{RelaxationRate0}
\Gamma_i^{(0)}=\frac{\sigma_i^{1/2}}{\tau_s\sqrt{\pi}}
\left(1-h^2\right)\left[\left(1+h\right)\,e^{-\sigma_i\left(1+h\right)^2}
+\left(1-h\right)\,e^{-\sigma_i\left(1-h\right)^2}\right]\equiv 
\frac{2\sigma_i^{1/2}}{\tau_s\sqrt{\pi}}\times \Upsilon(\sigma_i,h),
\end{equation}
\end{widetext}
with $\tau _{s}=\left(\lambda\gamma H_a\right) ^{-1}$.
In the high-energy barrier approximation, $\sigma \gg 1,h=x/2\sigma \ll 1$ the function $F_i$ reads \cite{garetal99pre}
\begin{equation}\label{Fiexpanded}
F_i\simeq 1-\frac{5}{4\lambda ^{2}}\frac{1}{\sigma _{i}},
\end{equation}
where $\lambda$ is the Landau-Lifshitz damping parameter.

One should note that the relaxation rate in Eq.~(\ref{RelaxationRate0}) applies to the case of a magnetic field applied along the anisotropy easy axis, and can then be rigorously used only for a textured assembly, i.e., with all anisotropy axes parallel to the applied field. For an assembly with randomly distributed easy axes, one should use the (cumbersome) expression of the relaxation rate in an oblique field \cite{cofetal95prb, coffeyetal98prb}.
In the present calculation we ignore this effect and use expression (\ref{RelaxationRate0}) for all moments in the assembly and then average over the direction of the anisotropy axes [see discussion in section \ref{sec:discussion}].
The same approximation was used in Ref.~\onlinecite{luisetal04jpcm}, while the calculations of  Ref.~\onlinecite{jongar01epl} of the ac susceptibility did not require a finite field and hence expression (\ref{RelaxationRate0}) was used at $h=0$.

The next step consists in substituting for $\mathbf{\xi}_i$ in Eq.~(\ref{jongar_rr}) the expression given by Eq.~(\ref{eff_field}) averaged over the spin and anisotropy orientations.
Averaging over the spin orientations yields [see appendix \ref{app:SpinAverage}]
\begin{equation}
\left\{ \begin{array}{l}
\left\langle \zeta_{i,\parallel}^{2}\right\rangle_{0}=\frac{\left(\xi_{d}n_{i}\right)^{2}}{3}\Theta_{i},\\
\\\left\langle \zeta_{i,\perp}^{2}\right\rangle _{0}=\frac{\left(\xi_{d}n_{i}\right)^{2}}{3}\Lambda_{i},\end{array}\right.\label{eq:final_effective_field}
\end{equation}
 where [see Eq.~(\ref{final_effective_field_comp}) et seq.]
\begin{eqnarray*}
\Theta_i &\equiv&
\sum\limits_j\,n_j^2\left[\left( 1-S_{j2}\right) \left( \mathbf{e}_{i}\cdot \mathcal{D}
_{ij}\cdot \mathcal{D}_{ij}\cdot \mathbf{e}_{i}\right) +3S_{j2}\Omega
_{ij}^{2}\right],\\
\Lambda_i &\equiv& \sum\limits_{j}n_{j}^{2}\left[ \frac{6}{
r_{ij}^{6}}+\frac{3}{r_{ij}^{3}}S_{j2}\Omega _{ij}\right]-\Theta_i, \\
\Omega_{ij} &\equiv& \mathbf{e}_{i}\cdot \mathcal{D}_{ij}\cdot \mathbf{e}_{j}.
\end{eqnarray*}

Therefore, the relaxation rate of a weakly interacting particle containing
$n_{i}$ Bohr magnetons, embedded in a polydisperse assembly, can be written as
\begin{equation}\label{DDIRR}
\Gamma_{i}\simeq\Gamma_{i}^{(0)}\left[1+\Xi_{i}\right].
\end{equation}
where we have collected the DDI contributions in
\begin{equation}\label{Xi}
\Xi_{i} = \frac{\left(\xi_{d}n_{i}\right)^{2}}{3}\frac{1}{2}\left(\Theta_{i} +\frac{F_i}{2}\Lambda_{i}\right).
\end{equation}

In the case of randomly distributed anisotropy easy axes one obtains [see appendix \ref{app:AnisotropyAverage}]
\begin{eqnarray*}
&&\overline{\mathbf{e}_{i}\cdot\mathcal{D}_{ij}\cdot\mathcal{D}_{ij}\cdot\mathbf{e}_{i}} = \frac{2}{r_{ij}^{6}},
\qquad\overline{\Theta_i}=\sum\limits_{j}\frac{2 n_j^2}{r_{ij}^{6}},\\
&&\overline{\Omega_{ij}} = 0, \qquad \overline{\Omega_{ij}^2} = \frac{2}{3}\frac{1}{r_{ij}^{6}},
\end{eqnarray*}
and
\begin{equation}\label{Xi_averaged}
\overline{\Xi_{i}} = \frac{\left(\xi_{d}n_{i}\right)^{2}}{3}\frac{1+F_i}{2}\,\overline{\Theta_i}.
\end{equation}

Now that we have the expression for the relaxation rate that includes the DDI contribution, we may study the effect of the latter on the critical volume $V_c$, or the corresponding number $n_{c}$ of Bohr magnetons, which is defined by the equation [see Eq.~(\ref{Susc5}) et seq.]
\begin{equation}\label{ncSPM}
\nu_{m}=\Gamma(n_{c}),
\end{equation}
where $\nu_{m}$ is the measuring frequency.
The problem then is to determine how $n_{c}$ (or $V_c$) changes in the presence of DDI, recalling that it is a function of temperature, field, and other experimental conditions such as $\nu_{m}$. For this we combine Eqs.~(\ref{RelaxationRate0}, \ref{DDIRR}, \ref{ncSPM}) to obtain
$$
\frac{\sqrt{\pi}\tau_{s}\nu_{m}}{2} \simeq \sqrt{\sigma}\Upsilon(\sigma,h)\left[1 + \overline{\Xi}\right].
$$
Now, using $\sigma_i = \sigma_0 n_i$ [see Eq. (\ref{dimless_params})] and dropping the index $i$, we rewrite this equation as
\begin{equation}\label{ncSPMwDDI}
\frac{1}{2}\ln n + \ln\Upsilon(\sigma_0, n, h) \simeq\ln\left(\frac{\sqrt{\pi}\tau_s\nu_{m}}{2\sqrt{\sigma_0}}\right) - \overline{\Xi}(\sigma_0, n).
\end{equation}

Since this equation has been derived in the case of weak DDI we may seek its solution $n_c$ as an expansion in terms of the DDI coefficient $\xi_d$ [see discussion of the validity of this perturbation in section \ref{sec:discussion}]. Indeed, inserting $n_c\simeq n_{c, 0} + \delta n_c$ in (\ref{ncSPMwDDI}) and expanding around $n_{c, 0}$, which is the solution of Eq.~(\ref{ncSPMwDDI}) without the DDI term $\overline{\Xi}$, we obtain the following expression for $\delta n_c$
\begin{equation}\label{Delta_nc}
\delta n_{c} = -\frac{n_{c, 0}\, \overline{\Xi}_0}{\frac{1}{2}  -  \sigma_0 n_{c, 0}\Phi(\sigma_c,h) + n_{c, 0}\overline{\Xi}^\prime_0},
\end{equation}
where $\sigma_c = \sigma_0 n_{c, 0}$ and
$$
\Phi(\sigma_c,h) \equiv \frac{\varphi_3(\sigma_c,h)}{\varphi_1(\sigma_c,h)}.
$$
with
$$
\varphi_n(\sigma_c,h) = (1+h)^n e^{-\sigma_c(1 + h)^2} + (1 - h)^n e^{-\sigma_c(1 - h)^2}.
$$
\begin{eqnarray*}
\overline{\Xi}_0 &\equiv& \overline{\Xi}(n_{c, 0}) = \overline{\Theta}_0\,\frac{\left(\xi_{d} n_{c, 0}\right)^{2}}{3}\left[1 - \frac{5}{8\lambda^{2}}\frac{1}{\sigma_{0}n_{c, 0}}\right]\\
\overline{\Xi}^\prime_0 &\equiv& \overline{\Xi}^\prime(n_{c, 0}) = 2 \overline{\Theta}_0\,\frac{\xi_d^{2}}{3}n_{c, 0}
\left[1-\frac{5}{16\lambda^{2}}\frac{1}{\sigma_{0}n_{c, 0}}\right].
\end{eqnarray*}
%
\begin{figure}[floatfix]
\includegraphics[width = 8.5cm]{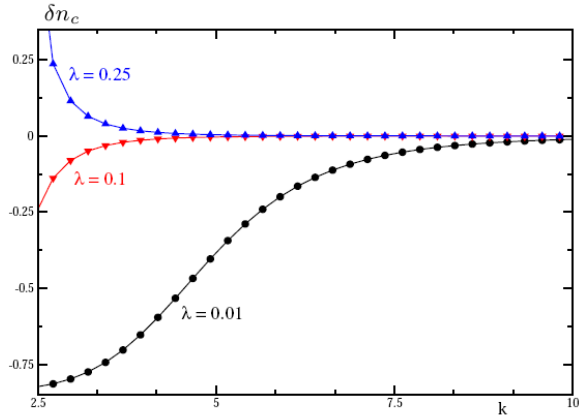}
\caption{\label{fig:Delta_nc_lambda}
(Color online) $\delta n_{c}$ versus the dimensionless parameter $k$ ($k = a/D_m$, $a$ being the inter-particle distance and $D_m$ the mean diameter) for the damping parameter $\lambda = 0.01, 0.1, 0.25$ and temperature $T=15\,\mathrm{K}$.}
\end{figure}
%
and
\begin{equation}\label{LatticeSum}
 \overline{\Theta}_0 = n_m^2\sum_j\frac{2}{r_{ij}^6} \equiv n_m^2\mathcal{R},
\end{equation}
where $\delta$ is the standard deviation of the volume distribution and $n_m = n_0\,e^{\delta^2}$ the mean number of Bohr magnetons corresponding to the mean volume of the assembly; $n_0$ being the number of Bohr magnetons contained in the volume $V_0$ [see Eq.~(\ref{LNdist})]. $\mathcal{R}\sim 16.8$ \cite{jongar01epl, jongar01prb} for a simple cubic lattice [see discussion in section \ref{sec:discussion}].

The function $\Phi(\sigma_c,h)$ decreases monotonically from $1$ to $0$ when $h$ varies from $0$ and tends to $1$, and is nearly independent of $\sigma$ especially for large $h$. This implies that when the applied field increases $\delta n_{c}$ increases (in absolute value) and thereby the effect of DDI is enhanced [see further discussion in the next section].
\section{\label{sec:discussion}Results and discussion}
In Fig.~\ref{fig:Delta_nc_lambda} we plot $\delta n_{c}$ as a function of the dimensionless parameter $k = a/D_m$ with varying damping parameter $\lambda$.
We see that for the relatively small values of $\lambda$, $\delta n_{c}$ becomes negative and decreases with the increasing intensity of DDI (or decreasing $k$). This means that the critical volume, separating the dominating populations of blocked and superparamagnetic particles, decreases in the presence of DDI.
This can also be seen from Eq.~(\ref{Fiexpanded}) upon noting that the function $F$, and thereby the contribution of the transverse component of the DDI field in Eq.~(\ref{jongar_rr}), changes sign upon varying the damping parameter.
A similar behavior was observed in Ref.~\cite{chuchan05jap} (see Fig.~2 therein) where the energy barrier distribution for coupled Co particles was computed (using a different approach) as a function of their concentration.  It was shown that the DDI induce an increment of the amount of small
barriers, responsible for faster decay. In our case, this is equivalent to the decrease of $V_c$ under the effect of DDI.
%
\begin{figure}[floatfix]
\includegraphics[width = 8.5cm]{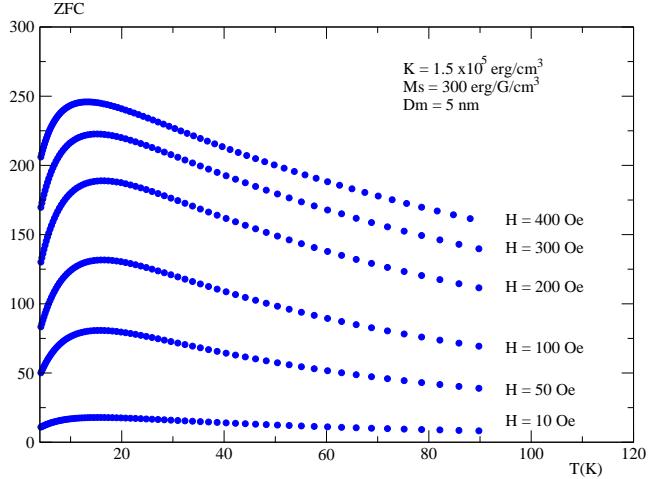}
\caption{\label{fig:ZFC}
(Color online) ZFC magnetization of a non-interacting assembly for various field values. $\delta=1$.}
\end{figure}
%

Obviously, all the curves (for different $\lambda$) tend to zero for large $k$ (absence of DDI). The limit between the negative and positive $\delta n_c$ is given by the equation $\lambda_\mathrm{limit}=\sqrt{5/(8\sigma_0 n_{c,0})}$, which implies that this limiting damping depends directly on temperature and on all other parameters via $n_{c,0}$.

Using the volume $V_c$ corresponding to $n_{c}\simeq n_{c,0} + \delta n_{c}$, with $n_{c,0}$ being the solution of Eq.~(\ref{ncSPMwDDI}) without the DDI term $\overline{\Xi}$ and $\delta n_{c}$ given by (\ref{Delta_nc}), we compute the ZFC magnetization according to Eq.~(\ref{MagSupMagBlock}).
In Fig.~\ref{fig:ZFC} we plot the ZFC magnetization thus obtained as a function of temperature for various values of the applied field of a polydisperse assembly of non-interacting maghemite nanoparticles with mean diameter  $D_m=5\,\mathrm{nm}$ and standard deviation $\delta=1.0$ and random anisotropy.
Apart from the obvious bell-like shape and the vertical shift of the maximum with the increasing field, we can see that the position of the maximum changes with the field in a non-monotonic way.

\begin{figure*}[floatfix]
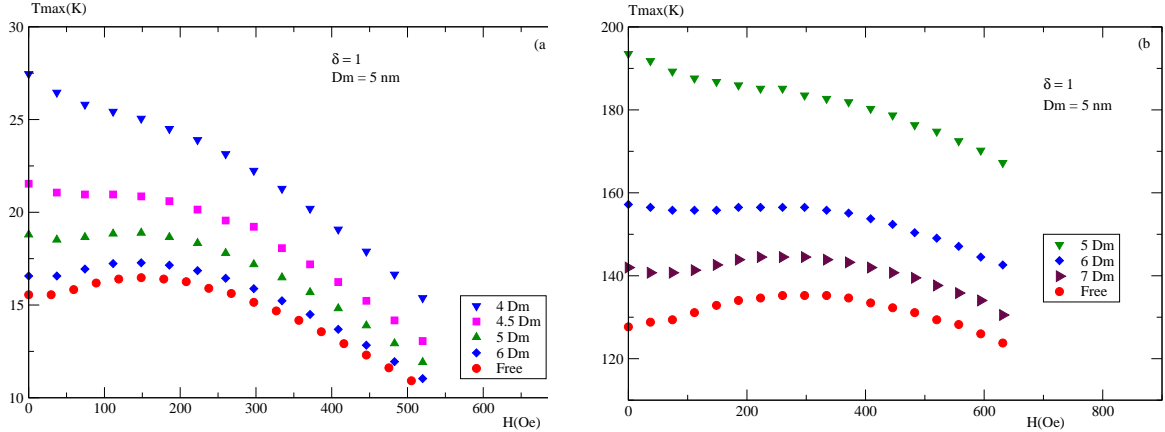

\includegraphics[width = 7.5cm]{TmaxH_Maghemite5nm_delta1p1_K1p5e5_Ms300_DDI.eps}\hspace{0.25cm}
\includegraphics[width = 7.5cm]{TmaxH_Cobalt5nm_delta1p1_K1p5e5_Ms1425_DDI.eps}
\caption{(Color online) The temperature $T_{\max}(H)$ as a function of the applied field for a) maghemite and b) cobalt particles for the same parameters $D_m$ and $\delta$ as in Fig.~\ref{fig:ZFC}. ``Free" stands for the non-interacting assembly and the other curves are for the interacting case with mean interparticle distance $a = k D_m$. The damping parameter is $\lambda = 0.01$.}
\label{fig:Tmax_DDI}
\end{figure*}

In Fig.~\ref{fig:Tmax_DDI} we plot the position of the maximum of the curves in Fig.~\ref{fig:ZFC} as a function of the applied field, i.e., $T_{\max}(H)$, for various values of the inter-particle (center-to-center) distance, for two substances.
%
%
First of all, we observe that indeed the effect of DDI is to change $T_{\max}(H)$ from a bell-like curve with a maximum to a monotonically decreasing function, and this compares well with the experimental results [see Fig. 1 of Ref.~\onlinecite{kacetal00jpcm}].
As was stressed in Ref.~\onlinecite{jongar01epl}, the effect of DDI is not merely to change the potential energyscape, as was argued in many previous publications \cite{doretal88jpc, mortro94prl, doretal97acp}, but also to introduce a transverse field that induces saddle points in the potential \cite{garetal99pre}.
This in turn makes the relaxation rate quite sensitive to the damping strength, and for relatively low damping, as is the case in Fig.~\ref{fig:Tmax_DDI} ($\lambda = 0.01$), the probability of switching increases.
In addition, if a magnetic field is added with increasing intensity, the energy barrier is lowered and the magnetic moments switch at lower temperatures.
This concomitant effect makes the relaxation time shorter and for a given observation time, the temperature $T_{\max}(H)$ increases.
This is of course compatible with the effect, discussed above, that DDI reduce the population of superparamagnetic particles in favor of the blocked ones and this leads to a larger $T_{\max}(H)$.
In summary, the critical volume $V_c$ (or equivalently $n_c$) given by Eq.~(\ref{ncSPMwDDI}) in fact corresponds to the ``critical" field at the maximum of $T_{\max}(H)$ which, in the presence of the DDI field, separates: i) the low-field regime: the population of blocked particles becomes more and more dominant, which leads to the increase of the ``average'' blocking temperature, and ii) the high-field regime: the population of superparamagnetic particles takes over thus leading to a decrease of the average blocking temperature.
Furthermore, increasing the intensity of the DDI field increases both its longitudinal and transverse components.
While the longitudinal component contributes to the increase of $T_{\max}$ the transverse component shifts its maximum towards low magnetic fields.

Finally, we recall that the work of Ref.~\onlinecite{kacetal00jpcm} showed that, in the absence of DDI, the maximum of $T_{\max}(H)$ can be explained by the nonlinear variation of the superparamagnetic contribution to the ZFC magnetization with the applied field. In the presence of DDI we see that the contribution of superparamagnetic particles is reduced in favor of blocked particles and this leads to the disappearance of the maximum.

The analytical expression (\ref{Delta_nc}) derived for $\delta n_c$ is valid, in principle, for $\xi_d n_m^2 \ll 1$ [see notation in Eq.~(\ref{dimless_params})], where $n_m$ is the mean number of Bohr magnetons in the assembly. This condition is equivalent to $\mid\delta n_c/n_{c,0}\mid \ll 1$ [see Eq.~(\ref{Delta_nc})] which leads to the condition on $k$

\begin{equation}\label{k_limit}
 k^3\gg\frac{2}{\sqrt{3}}\left(\frac{\mu_{0}}{4\pi}\right)\frac{\left(\mu_{B}^{2}/D_{m}^{3}\right)n_{c,0}}{k_{B}T}\times\sqrt{\left|\frac{\overline{\Theta}_{0}\, f_c}{\frac{1}{2}-\sigma_{c}\Phi_{c}}\right|},
\end{equation}
where
$$\sigma_c = \sigma_0\,n_{c,0},\quad f_c = 1-\frac{5}{16\lambda^{2}}\frac{1}{\sigma_{c}},\quad \Phi_{c} = \Phi(\sigma_c,h).$$
Then, using the physical parameters of Fig.~\ref{fig:Tmax_DDI} (left) with $D_m=5\,\mathrm{nm}, n_m\simeq 2117$, taking $T=20\,\mathrm{K}, H=100\,\mathrm{Oe}$ and computing $n_{c,0}$ from Eq.~(\ref{ncSPMwDDI}) without the DDI term, Eq.~(\ref{k_limit}) yields $k\gg 3$, which is a reasonable condition. Accordingly, in Fig.~\ref{fig:Tmax_DDI} we took $k \geq 4$.

Regarding the expression of the relaxation rate (\ref{RelaxationRate0}) a few remarks are in order. 
As we said earlier, rigorously, this expression applies to the case of a textured assembly with all easy axes parallel to the applied field. For random anisotropy, one should employ a numerical procedure for computing the relaxation rate with an oblique field, as is done in, e.g.,  Ref.~\onlinecite{cofetal95prb, coffeyetal98prb}.
However, we recall that the main objective of the present work was to i) understand the effect of DDI on the $T_{\max}(H)$ curve and ii) 
provide relatively simple (approximate) expressions including the DDI contribution. In addition, as the latter is only possible using perturbation theory which assumes weak DDI, a transverse magnetic field would be dominating and the subtle effect of the DDI field transverse component would not be easy to disentangle in a non ambiguous manner.
On the other hand, we have shown here that the disappearance of the maximum of $T_{\max}(H)$ is mainly due to the effect of the transverse component of the DDI field on the relaxation rate of the magnetic moments. One could then ask why the transverse component of the applied magnetic field does not play the same role in the case of a free assembly. 
The main reason is that the applied magnetic field has a static effect while the DDI provide a field that changes dynamically with temperature and other physical parameters related with the dynamics of the system. In this respect, we wish to make a connection with the work \cite{kac03epl} where the effect of exchange interaction on the relaxation rate of a two-spin system was (semi-analytically) investigated by the kinetic Langer's theory. It was shown that, in the weak coupling regime, when the first spin starts its switching process and arrives at the saddle point, the orientation of the second spin undergoes some fluctuations creating a small transverse field that increases the switching probability or the relaxation rate.

Furthermore, we would like to point out that in the present work we consider an assembly of nanoparticles placed at the sites of a regular (simple cubic) lattice, with varying inter-particle distances. The effect of changing the lattice structure or equivalently, the distribution of the vectors ${\bf r}_{ij}$ [see Eq.~(\ref{DDITensors}) et seq.], is to modify the lattice sum in Eq.~(\ref{LatticeSum}), and thereby to change $\delta n_c$ in Eq.~(\ref{Delta_nc}). In Ref.~\onlinecite{jongar01epl} the lattice sum ${\cal R}$ ($\propto\overline{\Theta}_0$, for a monodisperse assembly) takes the values $16.8, 14.5, 14.5$ for simple cubic, bcc, and fcc lattices, which leads to weaker DDI.
On the other hand, in many realistic samples the position of the particles on the hosting matrix is random, and for a given concentration, the precise variation of the lattice sum $\overline{\Theta}_0$ should depend on the particular form of the particles spatial distribution function, which may dramatically change in the presence of aggregates, chains, and the like.

We also found that changing the volume distribution $\delta$ width has only a quantitative effect on $T_{\max}(H)$, very much similar to the results presented in Fig.~4 of Ref.~\onlinecite{kacetal00jpcm} for a free assembly. Hence, in both interacting and non-interacting assemblies, we find that when the volume distribution becomes narrower, $T_{\max}$ decreases in magnitude and slightly flattens. 

In addition, changing from maghemite to cobalt particles, which is mainly equivalent (in the present approach) to changing the anisotropy constant by an order of magnitude, has a quantitative effect on the curve $T_{\max}(H, \xi_d)$ curves but the qualitative features remain the same.
\section{Conclusion}
We have investigated the effect of (weak) dipolar interactions on the zero-field-cooled magnetization by computing their contribution to the longitudinal relaxation time.
We have shown that the effect of the dipolar interactions is to lower the critical volume of the assembly which separates the dominating populations of blocked and superparamagnetic particles. More precisely, it is demonstrated that the maximum of $T_{\max}(H)$ shifts towards low values of the applied field as the intensity of the dipolar interactions, or equivalently the sample concentration, increases. This result is in good qualitative agreement with experiments on both maghemite and cobalt nanoparticles.
We finally emphasize the important role played by damping in the presence of a transverse field provided here by the dipolar interaction.
\newpage\appendix
\section{\label{app:SpinAverage}Spin averages}

We compute the average of the square of the longitudinal and transverse components with respect to the
local easy axis $\mathbf{e}_i$ of the effective field $\mathbf{\zeta}_i$ [see Eq.~(\ref{jongar_rr})] comprising both the applied magnetic and the DDI fields with the condition $\vert\zeta\vert\ll 1$.
The magnetic field is included here only for completeness and is dropped in the final expressions obtained in these appendices. In fact, the calculation of the ZFC magnetization and thereby that of $T_{\max}(H)$ requires the full range of this field, and for this reason in the calculation of $T_{\max}(H)$ the magnetic field is included exactly in the relaxation rate [see Eq.~(\ref{RelaxationRate0})]. 
Nevertheless, the expressions of the longitudinal and transverse components of the effective field $\mathbf{\zeta}_i$ obtained here may be used in the range of small magnetic fields.

The effective local field $\mathbf{\zeta}_i$ reads
\begin{equation}\label{ZetaEffectiveField}
\mathbf{\zeta}_i = x_i\mathbf{e}_h + \xi_d n_{i}\sum_{j}\mathcal{D}_{ij}\cdot\mathbf{S}_j,
\end{equation}
where $\mathbf{S}_i \equiv n_i \mathbf{s}_i$.
Its longitudinal component is then defined as
\begin{equation} \label{eff_field_parallel}
\zeta _{i,\parallel}=\mathbf{\zeta }_{i}\cdot \mathbf{e}_{i}=x_{i}(\mathbf{e%
}_{h}\cdot \mathbf{e}_{i})+\xi _{d}n_{i}\sum_{j}\mathbf{e}_{i}\cdot \mathcal{%
D}_{ij}\cdot \mathbf{S}_{j}.
\end{equation}
and thus
\begin{widetext}
\begin{eqnarray*}
\left\langle \zeta _{i,\parallel }^{2}\right\rangle _{0} &=& x_{i}^{2}\left(
\mathbf{e}_{h}\cdot \mathbf{e}_{i}\right) ^{2} + 2x_{i}\xi _{d}n_{i}\left( 
\mathbf{e}_{h}\cdot \mathbf{e}_{i}\right) \sum_{j}\mathbf{e}_{i}\cdot 
\mathcal{D}_{ij}\cdot \left\langle \mathbf{S}_{j}\right\rangle _{0} \\
&&+ \left(
\xi _{d}n_{i}\right) ^{2}\sum_{j}\sum_{k}\left\langle \left( \mathbf{e}
_{i}\cdot \mathcal{D}_{ij}\cdot \mathbf{S}_{j}\right) \left( \mathbf{e}
_{i}\cdot \mathcal{D}_{ik}\cdot \mathbf{S}_{k}\right) \right\rangle _{0} 
\end{eqnarray*}
\end{widetext}
where the average $\left\langle {}\right\rangle _{0}$ is defined with respect to the Gibbs probability distribution containing only the anisotropy term.

Now, using the following formulas
\begin{equation}\label{firstsecondaverages}
\left\langle s_{i}^{\alpha }\right\rangle _{0}=0,\quad \left\langle
s_{j}^{\alpha }s_{k}^{\beta }\right\rangle _{0}=\left[ \frac{1}{3}%
(1-S_{j2})\delta ^{\alpha \beta }+S_{j2}e_{j}^{\alpha }e_{j}^{\beta }\right]
\delta _{jk}
\end{equation}
with \cite{raiste97prb, jongar01prb}
$$
S_{il}(\sigma_i)\simeq
\displaystyle\left\lbrace
\begin{array}{ll}
\frac{(l-1)!!}{(2l+1)!!}(\frac{\sigma_i}{2})^{l/2}+\ldots, &\sigma_i \ll 1, \\ \\
1 - \frac{l(l+1)}{4\sigma_i}+\ldots, &\sigma_i \gg 1
\end{array}
\right.
$$
we obtain
\begin{widetext}
\begin{eqnarray}\label{zetaparallel}
\left\langle \zeta _{i,\parallel }^{2}\right\rangle _{0} &=&x_{i}^{2}\left( 
\mathbf{e}_{h}\cdot \mathbf{e}_{i}\right) ^{2} + 2x_{i}\xi _{d}n_{i}\left( 
\mathbf{e}_{h}\cdot \mathbf{e}_{i}\right) \sum_{j}\mathbf{e}_{i}\cdot 
\mathcal{D}_{ij}\cdot \left\langle \mathbf{S}_{j}\right\rangle _{0} \\ 
&&+ \left(
\xi _{d}n_{i}\right)^{2}\sum_{j}\sum_{k}\left\langle \left( \mathbf{e}
_{i}\cdot \mathcal{D}_{ij}\cdot \mathbf{S}_{j}\right) \left( \mathbf{e}
_{i}\cdot \mathcal{D}_{ik}\cdot \mathbf{S}_{k}\right) \right\rangle _{0} 
\nonumber \\
& = &x_{i}^{2}\left( \mathbf{e}_{h}\cdot \mathbf{e}_{i}\right) ^{2}+\frac{
\left( \xi _{d}n_{i}\right) ^{2}}{3}\sum_{j}n_{j}^{2}\left[ \left(
1-S_{j2}\right) \left( \mathbf{e}_{i}\cdot \mathcal{D}_{ij}\cdot \mathcal{D}
_{ij}\cdot \mathbf{e}_{i}\right) +3S_{j2}\left( \mathbf{e}_{i}\cdot \mathcal{
D}_{ij}\cdot \mathbf{e}_{j}\right) ^{2}\right], \nonumber
\end{eqnarray}
\end{widetext}

The transverse field is given by $\left\langle \zeta _{i,\perp }^{2}\right\rangle_{0}=\left\langle \zeta _{i}^{2}\right\rangle _{0}-\left\langle \zeta
_{i,\parallel }^{2}\right\rangle _{0}$,
with
\[
\left\langle \zeta _{i}^{2}\right\rangle _{0}=x_{i}^{2}+\left( \xi
_{d}n_{i}\right) ^{2}\sum_{j}n_{j}^{2}\left\langle \left( \mathcal{D}_{ij}%
\mathbf{s}_{j}\right) ^{2}\right\rangle _{0} 
\]
\begin{widetext}
\begin{eqnarray}  \label{eff_field_sqr_average}
\left\langle \zeta _{i,\perp }^{2}\right\rangle _{0} &=&x_{i}^{2}\left(1-\left( \mathbf{e}_{h}\cdot \mathbf{e}_{i}\right) ^{2}\right) +\left( \xi_{d}n_{i}\right) ^{2}\sum_{j}n_{j}^{2}\left\langle \left( \mathcal{D}_{ij}
\mathbf{s}_{j}\right) ^{2}\right\rangle _{0} \\
&-&\frac{\left( \xi _{d}n_{i}\right) ^{2}}{3}\sum_{j}n_{j}^{2}\left[ \left(
1-S_{j2}\right) \left( \mathbf{e}_{i}\cdot \mathcal{D}_{ij}\cdot \mathcal{D}
_{ij}\cdot \mathbf{e}_{i}\right) +3S_{j2}\left( \mathbf{e}_{i}\cdot \mathcal{
D}_{ij}\cdot \mathbf{e}_{j}\right) ^{2}\right]  \nonumber \\
&=& x_{i}^{2}\left( 1-\left( \mathbf{e}_{h}\cdot \mathbf{e}_{i}\right)^{2}\right) \\
&&+ \frac{\left( \xi _{d}n_{i}\right) ^{2}}{3}\sum_{j}n_{j}^{2} 
\left[
\begin{array}{ll}
3\left\langle \left( \mathcal{D}_{ij}\mathbf{s}_{j}\right) ^{2}\right\rangle
_{0} - \left( 1-S_{j2}\right) \left( \mathbf{e}_{i}\cdot \mathcal{D}_{ij}\cdot \mathcal{D}_{ij}\cdot \mathbf{e}_{i}\right) \\
- 3S_{j2}\left( \mathbf{e}_{i}\cdot \mathcal{D}_{ij}\cdot \mathbf{e} _{j}\right) ^{2}
\end{array}
\right] . 
\nonumber
\end{eqnarray}
\end{widetext}

Next, using Eq. (\ref{firstsecondaverages}), we compute the first term in
the square brackets as
\begin{widetext}
\begin{eqnarray*}
\left\langle \left( \mathcal{D}_{ij}\mathbf{s}_{j}\right) ^{2}\right\rangle
_{0} &=&\left\langle \left( \mathcal{D}_{ij}\mathbf{s}_{j}\right) \left( 
\mathcal{D}_{ij}\mathbf{s}_{j}\right) \right\rangle _{0}=\sum\limits_{\alpha
\beta \gamma }\mathcal{D}_{ij}^{\alpha \beta }\mathcal{D}_{ij}^{\alpha
\gamma }\left\langle s_{j}^{\beta }s_{j}^{\gamma }\right\rangle_{0}\\
&=& \frac{1}{3}\left[ (1-S_{j2})\sum\limits_{\alpha \beta }\mathcal{D}
_{ij}^{\alpha \beta }\mathcal{D}_{ij}^{\alpha \beta
}+3S_{j2}\sum\limits_{\alpha \beta \gamma }\left( e_{j}^{\beta }\mathcal{D}
_{ij}^{\alpha \beta }\right) \left( \mathcal{D}_{ij}^{\alpha \gamma
}e_{j}^{\gamma }\right) \right]
\end{eqnarray*}
\end{widetext}

Let us now compute these two terms.
\begin{widetext}
\begin{eqnarray*}
\sum\limits_{\alpha \beta }\mathcal{D}_{ij}^{\alpha \beta }\mathcal{D}_{ij}^{\alpha \beta } &=&\sum\limits_{\alpha \beta }\frac{3e_{ij}^{\alpha
}e_{ij}^{\beta }-\delta ^{\alpha \beta }}{r_{ij}^{3}}\times \frac{3e_{ij}^{\alpha }e_{ij}^{\beta }-\delta ^{\alpha \beta }}{r_{ij}^{3}}\\
&=&\frac{9}{r_{ij}^{6}}\sum\limits_{\alpha \beta }e_{ij}^{\alpha}e_{ij}^{\beta }e_{ij}^{\alpha }e_{ij}^{\beta }+\frac{1}{r_{ij}^{6}}\sum\limits_{\alpha \beta }\delta ^{\alpha \beta }\delta ^{\alpha \beta }-\frac{6}{r_{ij}^{6}}\sum\limits_{\alpha }e_{ij}^{\alpha }e_{ij}^{\alpha} = \frac{6}{r_{ij}^{6}},
\end{eqnarray*}
\end{widetext}

Next,
\begin{widetext}
\begin{eqnarray*}
\sum\limits_{\alpha }\mathcal{D}_{ij}^{\alpha \beta }\mathcal{D}%
_{ij}^{\alpha \gamma } &=&\frac{1}{r_{ij}^{6}}\sum\limits_{\alpha }\left[
3e_{ij}^{\alpha }e_{ij}^{\beta }-\delta ^{\alpha \beta }\right] \left[
3e_{ij}^{\alpha }e_{ij}^{\gamma }-\delta ^{\alpha \gamma }\right] \\
&=&\frac{1}{r_{ij}^{6}}\left[ 9e_{ij}^{\beta }e_{ij}^{\gamma
}\sum\limits_{\alpha }\left( e_{ij}^{\alpha }e_{ij}^{\alpha }\right)
-3e_{ij}^{\beta }\sum\limits_{\alpha }e_{ij}^{\alpha }\delta ^{\alpha \gamma
}-3e_{ij}^{\gamma }\sum\limits_{\alpha }e_{ij}^{\alpha }\delta ^{\alpha
\beta }+\sum\limits_{\alpha }\delta ^{\alpha \beta }\delta ^{\alpha \gamma
}\right] \\
&=&\frac{1}{r_{ij}^{6}}\left[ 3e_{ij}^{\beta
}e_{ij}^{\gamma }-\delta ^{\beta \gamma }\right] +\frac{2}{r_{ij}^{6}}\delta
^{\beta \gamma } = \frac{1}{r_{ij}^{3}}\mathcal{D}_{ij}^{\beta \gamma }+\frac{2}{r_{ij}^{6}}
\delta ^{\beta \gamma }
\end{eqnarray*}
\end{widetext}

then
\begin{widetext}
\begin{eqnarray*}
\sum\limits_{\alpha \beta \gamma }\left( e_{j}^{\beta }\mathcal{D}_{ij}^{\alpha \beta }\right) \left( \mathcal{D}_{ij}^{\alpha \gamma}e_{j}^{\gamma }\right) &=&\sum\limits_{\alpha \beta \gamma }e_{j}^{\beta}e_{j}^{\gamma }\mathcal{D}_{ij}^{\alpha \beta }\mathcal{D}_{ij}^{\alpha\gamma }=\frac{1}{r_{ij}^{3}}\sum\limits_{\beta \gamma }e_{j}^{\beta }\left[\frac{1}{r_{ij}^{3}}\mathcal{D}_{ij}^{\beta \gamma }+\frac{2}{r_{ij}^{6}}\delta ^{\beta \gamma }\right] e_{j}^{\gamma}\\
&=& \frac{1}{r_{ij}^{3}}\mathbf{e}_{j}\cdot \mathcal{D}_{ij}\cdot \mathbf{e}_{j}+\frac{2}{r_{ij}^{6}}
\end{eqnarray*}
\end{widetext}

Recapitulating, we have 
\begin{widetext}
\begin{eqnarray*}
\left\langle \left( \mathcal{D}_{ij}\mathbf{s}_{j}\right) ^{2}\right\rangle
_{0} &=& \frac{1}{3}\left[ \frac{6}{r_{ij}^{6}}-S_{j2}\frac{6}{r_{ij}^{6}}+\frac{
1}{r_{ij}^{3}}3S_{j2}\mathbf{e}_{j}\cdot \mathcal{D}_{ij}\cdot \mathbf{e}_{j}+\frac{6}{r_{ij}^{6}}S_{j2}\right]\\
&=& \frac{1}{3}\left[ \frac{6}{r_{ij}^{6}
}+\frac{1}{r_{ij}^{3}}3S_{j2}\left( \mathbf{e}_{j}\cdot \mathcal{D}
_{ij}\cdot \mathbf{e}_{j}\right) \right] . 
\end{eqnarray*}
\end{widetext}

Therefore, inserting all results back into Eq. (\ref{eff_field_sqr_average}), we finally obtain 
%
\begin{eqnarray*}\label{final_zeta_perp}
\left\langle \zeta _{i,\perp }^{2}\right\rangle _{0} &=& x_{i}^{2}\left(1-\left( \mathbf{e}_{h}\cdot \mathbf{e}_{i}\right)^{2}\right) \\
&& + \frac{\left(\xi _{d}n_{i}\right)^{2}}{3}\sum_{j}n_{j}^{2}\left[
\begin{array}{ll}
 \frac{6}{
r_{ij}^{6}} + \frac{3}{r_{ij}^{3}}S_{j2}\left(\mathbf{e}_{j}\cdot \mathcal{D}_{ij}\cdot \mathbf{e}_{j}\right) \\
- \left( 1-S_{j2}\right) \left( \mathbf{e}_{i}\cdot \mathcal{D}_{ij}\cdot \mathcal{D}_{ij}\cdot \mathbf{e}_{i}\right)-3S_{j2}\left( \mathbf{e}_{i}\cdot \mathcal{D}_{ij}\cdot \mathbf{e}_{j}\right) ^{2}
\end{array}
\right].
\end{eqnarray*}
%

Gathering the results for both longitudinal and transverse components for
the effective field, we write
\begin{widetext} 
\begin{equation}
\left\{ 
\begin{array}{l}
\left\langle \zeta _{i,\parallel }^{2}\right\rangle _{0}=x_{i}^{2}\left( 
\mathbf{e}_{h}\cdot \mathbf{e}_{i}\right) ^{2}+\frac{\left( \xi
_{d}n_{i}\right) ^{2}}{3}\sum\limits_{j}n_{j}^{2}\Theta_{ij}, \\
\\ 
\left\langle \zeta _{i,\perp }^{2}\right\rangle _{0}=x_{i}^{2}\left[
1-\left( \mathbf{e}_{h}\cdot \mathbf{e}_{i}\right) ^{2}\right]
+\frac{\left(\xi _{d}n_{i}\right) ^{2}}{3}\sum\limits_{j}n_{j}^{2}\left[ \frac{6}{r_{ij}^{6}}+\frac{3}{r_{ij}^{3}}S_{j2}\Omega _{ij}-\Theta_{ij}\right] ,
\end{array}
\right.  \label{final_effective_field_comp}
\end{equation}
\end{widetext}
where we have introduced the notation 
\begin{eqnarray*}
\Theta_{ij} &\equiv& \left( 1-S_{j2}\right) \left( \mathbf{e}_{i}\cdot \mathcal{D}
_{ij}\cdot \mathcal{D}_{ij}\cdot \mathbf{e}_{i}\right) +3S_{j2}\Omega
_{ij}^{2},\\
\Omega _{ij} &\equiv& \mathbf{e}_{i}\cdot \mathcal{D}
_{ij}\cdot \mathbf{e}_j. 
\end{eqnarray*}

\section{\label{app:AnisotropyAverage}Averaging over anisotropy}
The general expressions for the longitudinal and transversal fields can be simplified in some relevant situations. For a textured assembly (with parallel anisotropy axes) we set all the $\mathbf{e}_i$ parallel to $\mathbf{e}$. For a system with randomly distributed anisotropy axes one replaces expressions involving $f(\mathbf{e}_i)$ by integrals $\int d^{2}e\,f(\mathbf{e})\equiv \overline{f}$, and uses $\overline{(\mathbf{e}\cdot \mathbf{v}_{1})(\mathbf{e}\cdot \mathbf{v}_{2})}=\frac{1}{3}\mathbf{v}
_{1}\cdot \mathbf{v}_{2}$.

In the present work we only deal with random anisotropy. In this case we have 
\[
\int d^{2}e\,\left( \mathbf{e}_{h}\cdot \mathbf{e}_{i}\right) ^{2} = \frac{1}{3},
\]
so that for the longitudinal component we obtain
\begin{widetext}
\begin{eqnarray*}
\overline{\left\langle \zeta _{i,\parallel }^{2}\right\rangle_0}  &=& \frac{1}{3}x_{i}^{2} +
\frac{\left(\xi _{d}n_{i}\right)^{2}}{3}\sum\limits_j n_j^2\left[ \left( 1-S_{j2}\right) \overline{\mathbf{e}_{i}\cdot \mathcal{D}_{ij}\cdot \mathcal{D}_{ij}\cdot \mathbf{e}_{i}} + 3S_{j2}\overline{\left[ \mathbf{e}_{i}\cdot \mathcal{D}_{ij}\cdot \mathbf{e}_{j}\right] ^{2}}\right] \\
&=& \frac{1}{3}x_{i}^{2}
+ \frac{\left(\xi_d n_i\right)^2}{3}\sum\limits_{j}\frac{2 n_j^2}{r_{ij}^{6}}
\end{eqnarray*}
\end{widetext}
where we have used the averages 
\begin{eqnarray*}
&&\overline{\mathbf{e}_{i}\cdot \mathcal{D}_{ij}\cdot \mathcal{D}_{ij}\cdot 
\mathbf{e}_{i}} =\frac{1}{r_{ij}^{6}}\overline{3\left( \mathbf{e}_{i}\cdot
\mathbf{e}_{ij}\right) ^{2}+1}=\frac{2}{r_{ij}^{6}}, \\
&&\overline{\left[ \mathbf{e}_{i}\cdot \mathcal{D}_{ij}\cdot \mathbf{e}
_{j}\right] ^{2}} =\frac{1}{r_{ij}^{6}}\overline{3\left( \mathbf{e}_{j}\cdot \mathbf{e}_{ij}\right) ^{2}+1}=\frac{2}{3}\frac{1}{r_{ij}^{6}}, \\
&&\overline{\Theta_{ij}} =\left( 1-S_{j2}\right) \overline{\left( \mathbf{e}_{i}\cdot \mathcal{D}_{ij}\cdot \mathcal{D}_{ij}\cdot \mathbf{e}_{i}\right) }+3S_{j2}\overline{\Omega _{ij}^{2}} \\
&&\hspace{0.64cm}=\frac{2}{r_{ij}^{6}}\left( 1-S_{j2}\right) +3\frac{2}{3}\frac{1}{
r_{ij}^{6}}S_{j2}=\frac{2}{r_{ij}^{6}}
\end{eqnarray*}

For the transverse component, 
\begin{eqnarray*}
\overline{\left\langle \zeta _{i,\perp }^{2}\right\rangle _{0}}&=&\frac{2}{3}x_{i}^{2} + \frac{\left( \xi _{d}n_{i}\right) ^{2}}{3}\sum\limits_{j}n_{j}^{2}\frac{6}{r_{ij}^{6}}\\
&+& \frac{\left( \xi _{d}n_{i}\right) ^{2}}{3}\sum\limits_{j}n_{j}^{2}S_{j2}\frac{3}{r_{ij}^{3}}\overline{\Omega _{ij}}-\frac{\left( \xi _{d}n_{i}\right) ^{2}}{3}\sum\limits_{j}n_{j}^{2}\overline{\Theta_{ij}} \\
&=&\frac{2}{3}x_{i}^{2} + \frac{\left(\xi_d n_i\right)^2}{3}\sum\limits_{j}\frac{4 n_j^2}{r_{ij}^{6}}\\
&+& \frac{\left(\xi_d n_i\right)^2}{3}\sum\limits_{j}\frac{3 n_{j}^{2}}{r_{ij}^{3}}S_{j2}\overline{\Omega_{ij}}
\end{eqnarray*}
with
\begin{eqnarray*}
\overline{\Omega _{ij}} &=& \overline{\mathbf{e}_{i}\cdot \mathcal{D}_{ij}\cdot\mathbf{e}_{j}} \\
&=& \frac{1}{r_{ij}^{3}}\left[ 3\overline{\left( \mathbf{e}_{i}\cdot \mathbf{e}_{ij}\right) \left( \mathbf{e}_{ij}\cdot \mathbf{e}_{j}\right) }-\overline{\mathbf{e}_{i}\cdot \mathbf{e}_{j}}\right] = 0
\end{eqnarray*}
and thereby
$$
\overline{\left\langle \zeta _{i,\perp }^{2}\right\rangle _{0}}=\frac{2}{3}%
x_{i}^{2} +\frac{\left(\xi_d n_i\right)^2}{3}\sum\limits_{j}\frac{4 n_j^2}{r_{ij}^{6}}.
$$

Finally, for random anisotropy we have 
\begin{equation}
\left\{ 
\begin{array}{l}
\overline{\left\langle \zeta _{i,\parallel }^{2}\right\rangle _{0}}=\frac{1}{3}x_{i}^{2}
+ \frac{\left(\xi_d n_i\right)^2}{3}\sum\limits_{j}\frac{2 n_j^2}{r_{ij}^{6}},\\
\overline{\left\langle \zeta _{i,\perp }^{2}\right\rangle _{0}}=\frac{2}{3}x_{i}^{2} +\frac{\left(\xi_d n_i\right)^2}{3}\sum\limits_{j}\frac{4 n_j^2}{r_{ij}^{6}} = 2 \overline{\left\langle \zeta _{i,\parallel }^{2}\right\rangle _{0}}.
\end{array}
\right.  \label{randomzeta}
\end{equation}

\acknowledgements
We thank D.S. Schmool and O. Chubykalo-Fesenko for reading the manuscript and suggesting improvements.
%

%
\end{document}